\documentclass[preprint,prl]{revtex4}%

\usepackage{graphicx}
\usepackage{dcolumn}
\usepackage{bm}

\usepackage{hyperref}
\usepackage{amsmath}
\begin{document}


\title{Enhanced spontaneous emission into the mode of a cavity QED system}
\author{M. L. Terraciano, R. Olson Knell, D. L. Freimund, L. A. Orozco}
\address{Joint Quantum Institute, Dept. of Physics, University of Maryland, College Park, MD
20742-4111, U.S.A.}
\author{J. P. Clemens, P. R. Rice}
\address{Dept. of Physics, Miami University, Oxford, OH 45056,
U.S.A.}

\begin{abstract}
We study the light generated by spontaneous emission into a mode
of a cavity QED system under weak excitation of the orthogonally
polarized mode. Operating in the intermediate regime of cavity QED
with comparable coherent and decoherent coupling constants, we
find an enhancement of the emission into the undriven cavity mode
by more than a factor of 18.5 over that expected by the solid
angle subtended by the mode. A model that incorporates three
atomic levels and two polarization modes quantitatively explains
the observations.


\end{abstract}

\maketitle


Cavity QED has been identified as an environment to transfer
information and entanglement between matter and light qubits
\cite{cirac97,rice06}. Information inside the system must exit
through one of the two available channels: cavity decay, at a rate
$\kappa$, or spontaneous emission, at a rate $\gamma_{tot}$. These
decay mechanisms allow for the creation of quantum interconnects
and information protocols, since they enable information flow into
and out of the quantum system. Spontaneous emission has then a
dual role in cavity QED; it is a source of decoherence, but it is
also a useful way to extract information out of the system. The
detection of a spontaneously emitted photon in cavity QED is an
unambiguous probe of the state of the atomic part of the
atom-cavity system.

Work in the past has focused on the bad cavity limit where the
enhancement and suppression of spontaneous emission is easy to
identify (see for example the article by Hinds in
Ref.~\cite{berman94}). Other studies of spontaneous emission in
cavity QED include geometries that allow observation of the atoms
from the side \cite{childs96} and studies of the atomic
fluorescence into the mode of the cavity with the atoms driven by
a laser that propagates perpendicular to the cavity axis
\cite{zhu88,hennrich05}. This letter presents our investigations
of light generated by a spontaneous emission process into a mode
of a driven optical cavity in a system in the intermediate
coupling regime of cavity QED, where the dipole coupling between a
single atom  and the mode of the cavity, $g$, is comparable to the
dissipation rates $\kappa$ and $\gamma_{tot}$.

\begin{figure}[h]
\leavevmode \centering
\includegraphics[width=5in]{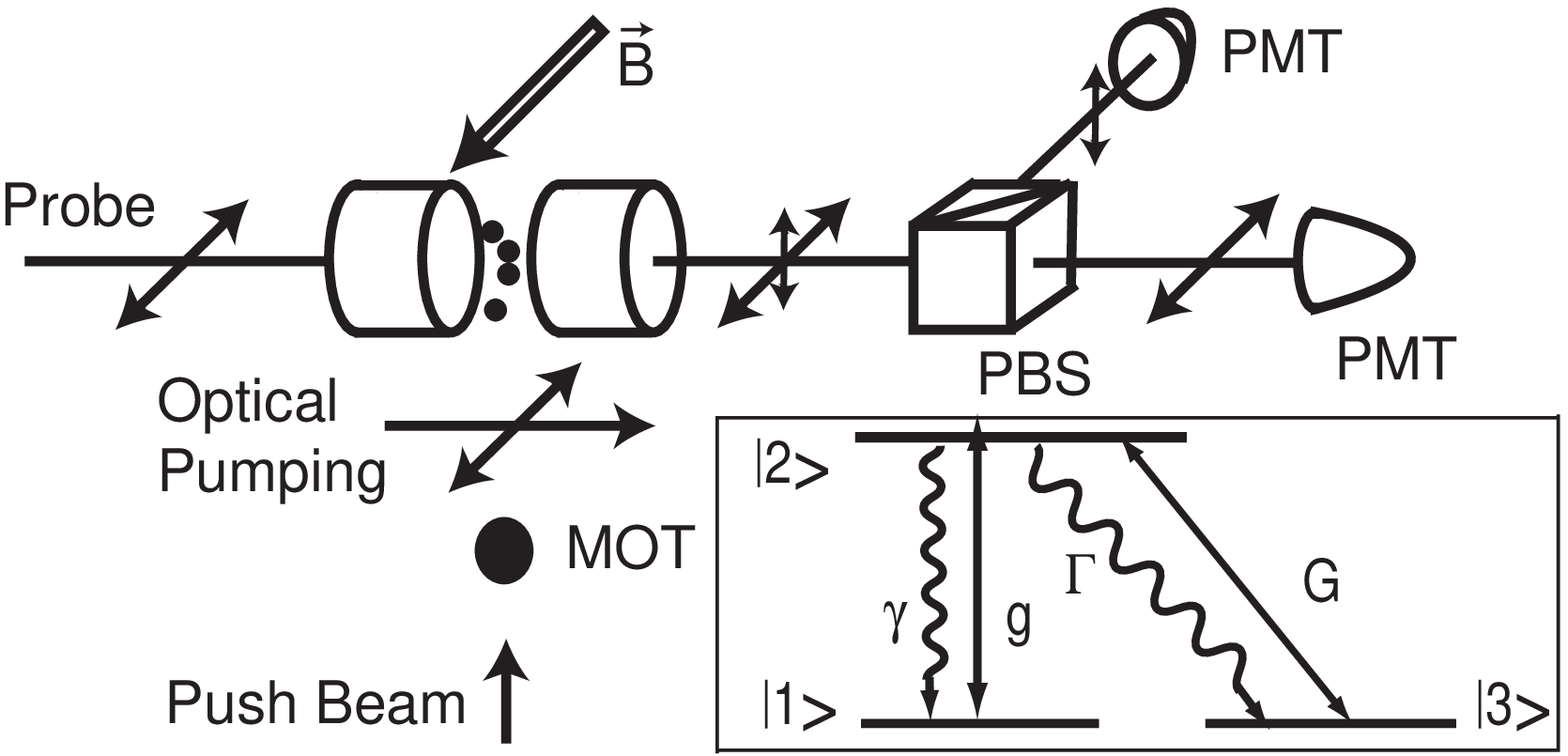}\caption{Schematic
of the experimental apparatus. A polarizer at the output separates
the two orthogonal linear polarizations. The inset shows the
energy level diagram used in the model.\label{figure2}}
\end{figure}

It is difficult to identify the origin of light exiting the mode
of the cavity in the intermediate regime of cavity QED. The light
can come from the drive, spontaneous emission, or stimulated
emission. We follow Birnbaum \emph{et al.} \cite{birnbaum05} by
using the polarization of the cavity emitted photon to gain
information on the origin of the photon and the state of the
atomic system.  Our cavity QED system consists of a high finesse
optical resonator where one or a few atoms interact with two
degenerate TEM$_{\text{00}}$ cavity modes with orthogonal linear
polarizations. We use the internal structure of the atoms to
inform us when a photon originates in a fluorescence event.
Instead of utilizing $^{85}$Rb atoms in their stretched states
($m_F=F$ with $\Delta m=1$) to form a closed two-level system when
driven with circularly polarized light, we prepare the atoms into
the $m_F=0$ ground state and drive the optical transition with
$\pi$ polarization ($\Delta m=0$). Next, we look at the light
emitted out of the cavity separating it into the two linear
polarizations, one parallel to the drive and the other orthogonal
to the drive (see Fig.~\ref{figure2} for a schematic of the
apparatus). The presence of any light of orthogonal polarization
signals that it comes originally from a spontaneous emission event
of an atom that decays emitting circularly polarized light with
$\Delta m = \pm 1$.

A model for the atoms that captures the essential physics of the
system needs to consider more than two atomic levels as sketched
in the inset of Fig.~\ref{figure2}. The modes of the cavity couple
to a single atom with coupling constants $g$ and $G$ that depend
on the electric dipole moment of the transition, with $G=g/\eta$.
The factor ($\eta$) takes into account the Clebsh Gordon
coefficients of the electric dipole moment that joins the two
levels, as well as any other factors that affect the coupling. The
inset shows the relevant energy level diagram. The respective
decay rates $\gamma$ and $\Gamma$ are also related by
$\Gamma=\gamma/\eta$. The system is driven weakly on-axis by a
classically horizontally polarized field $\varepsilon/\kappa$
normalized to photon flux units.

We consider $N$ three-level atoms fixed with degenerate ground
states maximally coupled to a two mode cavity with orthogonal
linear polarizations. The effective Hamiltonian for the
between-jump evolution in quantum trajectory theory
\cite{carmichael93book} is
\begin{eqnarray}
H &=& \frac{\varepsilon}{\kappa}\left(a-a^\dagger\right)+i\hbar g\left(a\sum_{i=1}^N |2\rangle_i\langle 1| - a^\dagger\sum_{i=1}^N |1\rangle_i\langle 2|\right) \nonumber \\
&&+ i\hbar G\left(c\sum_{i=1}^N |2\rangle_i\langle 3|- c^\dagger\sum_{i=1}^N |3\rangle_i\langle 2|\right)   \nonumber \\
&&- i\hbar\kappa\left(a^\dagger a + b^\dagger b\right) -
i\hbar\frac{\gamma+\Gamma}{2}\sum_{i=1}^N |2\rangle_i\langle 2|
\label{hamiltonian}
\end{eqnarray}
where $a$ is the annihilation operator for the driven horizontally
polarized cavity mode, $b$ is the annihilation operator for the
vertically polarized cavity mode, and $c=(a+ib)/\sqrt{2}$.

The state of the system to first order in $\varepsilon/\kappa$ is:
\begin{eqnarray}
|\Psi\rangle&=& |0000\rangle + c_{0010}|0010\rangle +
c_{1000}|1000\rangle \nonumber\\
 && +c_{0100}|0100\rangle +
c_{1001}|1001\rangle + c_{0101}|0101\rangle, \label{psi}
\end{eqnarray}
the labels in the probability amplitudes,
$|n_a,n_b,n_2,n_3\rangle$, denote the photon number for the $a$
mode, $b$ mode, number of atoms in $|2\rangle$, number of atoms in
$|3\rangle$, and with the number of atoms in $|1\rangle$,
$n_1=N-n_2-n_3$. The atomic portion of the state is a collective
state referring to $N$ atoms symmetric with respect to the
exchange of any pair of atoms.

We find the equations of motion for the coefficients using the
Hamiltonian and solve them in steady state starting with $N$ atoms
in $|1\rangle$. The driven mode transmission is proportional to
the steady state solution of $|c_{1000}|_{ss}^2$, while the
undriven mode transmission is proportional to $|c_{0101}|_{ss}^2$.

We use two dimensionless numbers to characterize the influence of
an atom in the system: $C_1=g^2/\kappa(\gamma+\Gamma)$, and
$\widetilde{C}_1=G^2/\kappa((\gamma+\Gamma))$. The transmitted
intensities normalized to the empty cavity intensity of the driven
mode are:
\begin{eqnarray}
T_{d}&=&\frac{1}{(1+2C_{1}N/(1+2\widetilde{C}_{1}))^2}; \label{td}\\
T_{u}&=&\left(\frac{2\widetilde{C}_1}{1+2\widetilde{C}_1}\right)\left(\frac{C_1N/(1+2\widetilde{C}_1)}
{\left(1+2C_1N/(1+2\widetilde{C}_1)\right)^2}\right). \label{tu}\
\end{eqnarray}
The first factor in Eq.~\ref{tu} is the ratio of enhanced
spontaneous emission into the orthogonal mode to the total
spontaneous emission, also known as the beta factor in laser
theory \cite{rice88,clemens04}. The second factor is the
probability of having an atom in the excited state under low
excitation. The dependence on $N$ in the theory, which assumes
very weak excitation, shows that the absorption of a photon from
the driven mode takes the system into a collective state with one
excited atom among $N$. When the photon is emitted into the
undriven mode, the transition is not collective and does not grow
with $\sqrt{N}$. The individual atoms do not return to the same
original state where they started.

The apparatus (see Fig. \ref{figure2}) consists of two main
components: The source of atoms and the cavity. A titanium
sapphire laser (Ti:Sapph) provides most of the light needed for
the experiment at 780 nm. The laser frequency is locked using a
Pound-Drever-Hall (PDH) technique on saturation spectroscopy of
$^{85}$Rb.

A rubidium dispenser delivers Rb vapor to a magneto-optical trap
(MOT) in a glass cell 20 cm below a cubic chamber that houses the
cavity. The glass cell has a silane coating to decrease the
sticking of Rb to the walls and maximize the capture efficiency of
the MOT \cite{aubin03a}. Each of the six 30 mW beams of the MOT
has a 1/e (power) diameter of 20 mm. A second laser repumps the
atoms that fall out of the cycling transition in the trap. A pair
of anti-Helmholtz coils generates a magnetic field gradient of 6
G/cm and three sets of independent coils zero the magnetic field
at the trapping region.

The cavity defines a TEM$_{00}$ mode with two 7 mm diameter
mirrors with different transmission coefficients. The input
transmission (15~ppm) is smaller than the output (250~ppm) to
ensure that most of the signal escapes from the cavity on the
detector side. The separation between the mirrors is 2.2 mm so the
coupling coefficient between the driven mode and the $\pi$ dipole
transition of Rb is $g/2\pi=1.5$ MHz. The finesse of the cavity is
${\cal F}\approx 11\,000$ with $\kappa/2\pi=3.2$ MHz. The mirrors
are glued directly to flat piezo-electric-transducers (PZT) for
controlling the length of the cavity. Our experimental system is
in the intermediate regime of cavity QED, where $g\approx ( \kappa
, \gamma_{tot}/2)$ with $(g, \kappa,
\gamma_{tot}/2)/2\pi=(1.5,3.2,3.0)$~MHz,
$\gamma_{tot}=\gamma+\Gamma$, giving $C_1$=0.12. The Clebsh Gordon
coefficients for the ($F=3,m_F=0 \rightarrow F'=4,m_F=0\pm1$) give
an optimal value for $\eta=\sqrt{8/3}$. The value would be larger
if some of the emitting atoms are not maximally coupled.

We stabilize the cavity length with a PDH technique using a 820 nm
laser. The frequency of this laser is locked to the stabilized 780
nm laser using a transfer cavity. We separate the two wavelengths
at the output of the physics cavity with a grating, and use
appropriate interference filters to further ensure the separation
of the two colors. We launch the atoms from the MOT towards the
cavity with a pulsed near-resonant push beam from below and
prepare them with an optical pumping beam. The cavity drive is
resonant with the $D_2$ line between ($F=3,m_F=0 \rightarrow
F'=4,m_F=0$) states. The repetition rate sets the number of atoms
delivered to the cavity as more or less atoms accumulate in the
MOT during the waiting period. We take data by recording the
transmitted light in the two orthogonal linear polarizations for
resonant excitation. There is a slight non-degeneracy of the two
orthogonal modes of less than 0.5 MHz (smaller than the full width
at half maximum of the transmission). The birefringence of the
cavity is less than $1\times 10^{-4}$ on its axis.

The geometry allows only $\pi$ excitations ($\Delta m=0$) and no
Faraday rotation of the light as the incoming polarization is
aligned with an external uniform magnetic field. The observed
light at the orthogonal polarization must come from spontaneous
emission. This light is emitted into the well defined spatial mode
of the cavity, facilitating its detection. The weak input field
drive $(\varepsilon/\kappa)$ is polarized horizontally to better
than $1 \times 10^{-5}$ and aligned to the magnetic field to
better than $\pm 4$ degrees. The output of the cavity is split
into two orthogonal polarization beams with a Glan-Laser
polarizer. Small changes in the magnetic field alignment or in the
optical pumping of the atoms do not make qualitative changes in
the results and only minor quantitative ones.

As each launch of atoms (every 150 ms) traverses the cavity we
record the transmission of both polarizations (horizontal and
vertical) in a digital storage scope to average over 20 launches
of atoms. The individual atoms take about 10 $\mu$s to traverse
the mode of the cavity. The batch of atoms crosses the cavity mode
in about 500 $\mu$s. The temperature of the atoms is less than 0.5
mK.

\begin{figure}[h] \leavevmode \centering
\includegraphics[width=5in]{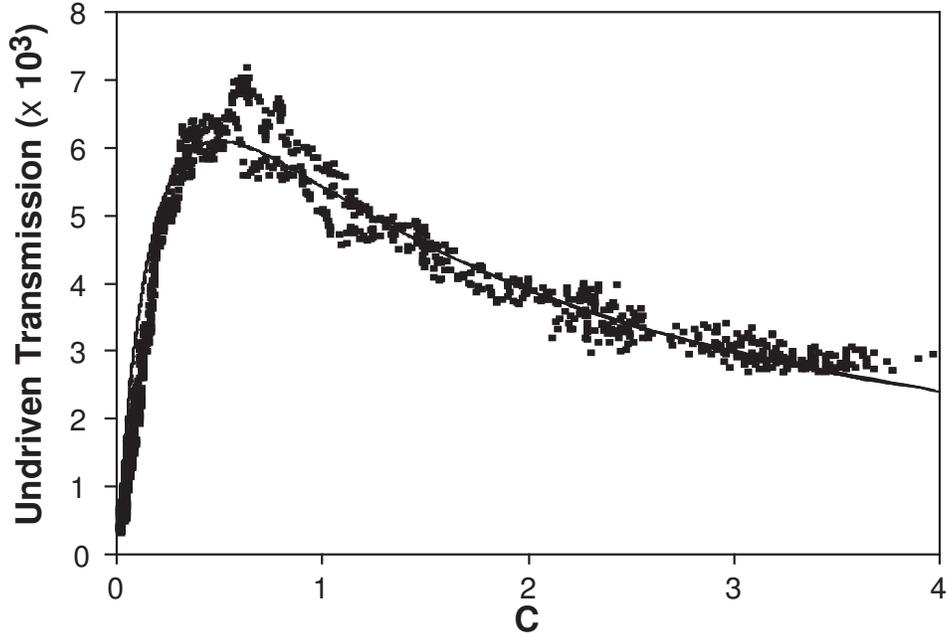}\caption{Variation of
the cavity transmission of the undriven mode (vertical
polarization) as a function of cooperativity $C$. The continuous
line is the prediction from the model. The vertical scale is
normalized to the empty cavity transmission ($C=0$) of the driven
mode. The range of the number of atoms is $0 \leq N \leq
36$.\label{t-N}}
\end{figure}

We parameterize the change in the transmission using the
cooperativity $C=C_1N/(1+2\widetilde{C}_1)$ of the driven mode. We
extract $C$ using the change in the normalized transmission of the
driven cavity mode (Eq~\ref{td}). The normalized transmission of
the driven mode, $T_d$, (horizontal polarization) decreases
monotonically as the number of atoms passing through increases.
The undriven mode (vertical polarization) shows a maximum on the
transmission, $T_u$, as the number of atoms increases (see
Fig.~\ref{t-N}). We extract a value of $\widetilde{C}_1=0.026 \pm
0.005$ with the only adjustable parameter in the model $\eta=2.1
\pm 0.4$. This value is consistent with the expected value. The
maximum of the transmission in the undriven mode happens at the
point where the atomic inversion is highest for a given drive:
$C=0.5$. This value of the cooperativity also coincides with the
point where the driven mode starts to show vacuum Rabi splitting
as a function of $N$.

The emitted light in the vertical mode comes from an atom in the
excited state. Its decay is through the coupling into the undriven
mode and it shows the enhanced spontaneous emission mediated by
$G$ (Eq.~\ref{tu}). The ratio between the undriven and driven
transmission, $T_u/T_d$ at the peak of the former ($C \approx 0.5$
in Fig.~\ref{t-N}) is $0.024 \pm 0.004$, where the uncertainty
includes the statistical error. The fraction subtended by the mode
of the cavity at the mirrors is $1.3\times 10^{-3}$, which would
be the fractional transmission, $T_u/T_d$ in the absence of
enhancement. Since we see a larger amount, $0.024$, we take the
ratio of this two numbers as the enhanced spontaneous emission
into the undriven mode: $18.5 \pm 3$, in agreement with the model.

 The labelling of the photons by polarization permits
us to identify an emission out of the cavity generated by an
excited atom spontaneous decay. The dependence of the light on the
number of atoms shows a maximum when the available weak drive
maximizes the atomic inversion.  The specific quantum dynamics of
the photon with orthogonal polarization remain to be explored, but
the intrinsic relation between the state of the atom and the
atomic polarization should allow exploration of atom-photon
correlations in cavity QED \cite{rice06}.

This work was supported by NSF and NIST. We thank H. J. Kimble and
H. J. Carmichael for their interest in this work.

\newpage

\newpage

\end{document}